\documentclass[twocolumn,amsmath,prb,amssymb,groupedaddress,showpacs,floatfix,society=aps]{revtex4-1}
\usepackage{amsmath}
\usepackage{todonotes}
\usepackage{subfig} 
\usepackage{bm}
\begin{document}
\title{AC-driven fractional quantum Hall systems: Uncovering unexpected features}
\author{Imen Taktak$^{1}$ and In\`es Safi$^{2}$}
 \affiliation{$1$: SPEC-CEA-Saclay, CNRS/UMR 3680, F-91191, Gif-sur-Yvette, France.\\
 $2$: Laboratoire de Physique des Solides, CNRS UMR 5802-University Paris-Saclay, France}
 \email{ines.saﬁ@universite-paris-saclay.fr}
\date{\today}
\begin{abstract}
We investigate the challenges of reaching the quantum regime and generating minimal excitations in the fractional quantum Hall effect (FQHE) under AC driving. Using the unifying non-equilibrium perturbative (UNEP) approach, we analyze weak backscattering through a quantum point contact (QPC). In both two-terminal geometries and the "anyon collider" setup, the lower bound on photo-assisted backscattering noise is set by the photo-assisted current rather than the DC noise predicted by Levitov’s theorem. This super-Poissonian character is confirmed within the Tomonaga-Luttinger liquid (TLL) framework, where Levitov’s theorem is violated, challenging the conventional interpretation of "photo-assisted" noise.  
In the two-terminal geometry, we show that when the QPC has a low scaling dimension, such as \( \delta = 1/3 \), two challenges arise: first, achieving the expected power-law behavior in the DC regime, and second, maintaining the AC quantum regime—where the drive frequency \( \omega_{ph} \) exceeds the temperature—when the DC voltage component is resonant with \( \omega_{ph} \).
The validity of the UNEP relations imposes a lower bound on temperature and a persisting equilibrium contribution to backscattering noise. This forces us to choose a higher $\delta=2/3$, for which we find that it is rather the photo-conductance—often overlooked— that might exhibit spikes at resonant DC voltages, providing a reliable probe of fractional charge. Moreover, we highlight that the zero temperature limit, frequently assumed in prior studies—including X.-G. Wen’s foundational work—is inappropriate in this context.  
Our findings extend beyond the FQHE to coherent conductors and Josephson or phase-slip junctions strongly coupled to an ohmic environment under AC bias, with broader implications for quantum transport and minimal excitation engineering.
\end{abstract}
\maketitle
\section{Introduction}
The Tomonaga-Luttinger (TLL) model\cite{schulz_revue} remains a central topic in both theoretical and experimental research. It serves as a paradigm for strongly correlated systems and provides a unified framework for studying various physical systems, such as edge states in the integer or fractional quantum Hall effect (FQHE)\cite{wen_review_FQHE_1992}, Josephson junctions\cite{grabert_ingold} and coherent conductors coupled to an Ohmic environment, which have been shown to offer quantum simulations of the TLL\cite{ines_saleur,zamoum_12,pierre_anne_boulat_universality}.
While experimental evidence for the TLL characteristic features, such as power-law behavior or the full Bethe-Ansatz solution\cite{fendley_prl_1}, has been inconclusive in the FQHE due to non-universal microscopic effects, it has been largely validated in quantum circuits exhibiting dynamical Coulomb blockade phenomena \cite{ines_pierre,pierre_PRX_2018,baranger_nature_environment_12}. Among the broad range of low-energy effective models proposed for the FQHE \cite{wen_FQHE_effective_theories_PRB_1990,jain_FQHE_compiste_fermions_PRL_1989}—some of which represent sophisticated generalizations of the TLL model—identifying a preferred model at a given fractional filling factor $\nu$ remains challenging due to a lack of clear theoretical or experimental evidence\cite{review_feldman}.\\
Despite this open issue on modelling the FQHE, one of its most intriguing features—the fractional charge $e^*$ of quasiparticles backscattered by an almost open quantum point contact (QPC)—has been experimentally observed\cite{saminad,heiblum_frac,kyrylo_note} using Poissonian non-equilibrium shot noise, initially predicted within the TLL model\cite{kane_fisher_noise,ines_resonance}, but persisting beyond the breakdown of the TLL description as demonstrated first in Ref.\cite{levitov_reznikov} and later extended through the unifying non-equilibrium perturbative (UNEP) approach \cite{ines_eugene,ines_cond_mat,ines_PRB_2019}. The latter applies to a much broader class of physical systems beyond tunneling junctions, without splitting the Hamiltonian, thus allowing for mutual interactions between opposite edges (see Eq.\eqref{eq:noise_DC_zero}). It unifies integer and fractional quantum Hall systems as well as normal and Josephson junctions (or their duals) in quantum circuits, regardless of whether the TLL model remains valid. Moreover, it can account for the absence of initial thermalization, leading to a non-equilibrium distribution where the DC noise becomes super-Poissonian \cite{ines_prb_long,ines_PRB_R_noise_2020,ines_photo_noise_PRB_2022} (see Eq.~\eqref{eq:dc_superpoisson}). This non-equilibrium distribution may arise from a temperature gradient or from QPC sources in an "anyon collider" \cite{fractional_statistics_gwendal_science_2020} (see Fig.~\ref{app:anyon_collider}). Notice that the term "anyon collider" is not strictly accurate, as the anyon fluxes injected by the source QPCs are too dilute to induce actual collisions at the central QPC.
Within the UNEP approach to time-dependent driven systems and finite frequency noise, we have derived many relations that offer alternative methods for determining the fractional charge, which are more robust than Poissonian noise \cite{ines_cond_mat,ines_eugene,ines_degiovanni_2016,ines_PRB_R_noise_2020,ines_photo_noise_PRB_2022}. Experimentally, such methods have been used to investigate fractional fillings $\nu$ beyond the Laughlin series, where the nature of the corresponding states has not gained unanimity\cite{glattli_photo_2018,ines_gwendal}. Addressing time-dependent voltages is also crucial for Hanbury-Brown and Twiss or Hong-Ou-Mandel setups designed to probe statistics \cite{buttiker_opera,Bocquillon_13_splitter_electrons_demand,levitons_glattli_review_2017,hbt,giuliano_HBT_FQHE_2016} or charge fractionnalization\cite{ines_epj,plasmon_ines_IQHE_Hong-Ou-Mandel_feve_Nature_2015_cite,sassetti_levitons_IQHE_2020}. These configurations fall within the scope of the UNEP approach, which has been used to interpret experimental findings in both integer and fractional quantum Hall effects within a unified framework\cite{Imen_thesis,glattli_imen_2022}.\\
The independence of the UNEP relations from the underlying microscopic model has been a significant advantage in all these experiments, where the measured DC current does not exhibit a clear signature of the TLL power-law behavior \cite{kyrylo_note}. For weak backscattering through an almost open QPC, the failure of the TLL description is actually advantageous for studying the effects of AC sources within the UNEP approach in the quantum regime (where the AC frequency satisfies \( \omega_{ph} \gg \omega_T = k_B T / \hbar \)). This is because the UNEP, when applied to the TLL model, necessitates several points of caution discussed in this paper and cannot reach the zero-temperature limit in the weak backscattering regime—a limit that has frequently but incorrectly been assumed in many prior theoretical studies. In particular, studies that proposed determining fractional charge through peaks at resonant DC voltage values (when $\omega_J=e^*V/\hbar=n\omega_{ph}$ with integer $n$) using photo-assisted current by X.-G. Wen\cite{wen_photo_PRB_91}, finite-frequency noise by C. Chamon {\it et al.} \cite{chamon_noise}, or photo-assisted noise by A. Cr\'epieux {\it et al.}\cite{crepieux_photo} did not adequately address their validity domain. We notice that such subtleties do not concern works on the tunneling of elementary electron charges, whether in the fractional\cite{sassetti_99_photo} or integer\cite{sassetti_levitons_IQHE_2020} quantum Hall regimes.
We also address the challenge of characterizing minimal excitations \cite{klich_levitov,glattli_levitons_nature_13} through backscattering noise in the FQHE. In particular, we show that an AC voltage superimposed on a DC voltage $V$ leads to a noise reduction over a broad range of $V$, a counterintuitive effect we have previously demonstrated in superconductor-insulator-superconductor junctions \cite{ines_photo_noise_PRB_2022}.  \\
This finding contradicts Levitov’s theorem (Eq.\eqref{eq:levitov's_theorem}), which applies only to linear conductors, making the term "photo-assisted" misleading—though we retain it for consistency with existing literature. Instead, we propose the expression \textbf{"coherent control of AC-driven quantum transport".} Within the validity domain of the UNEP relations, the appropriate generalization of Levitov’s theorem to nonlinear conductors is a super-Poissonian photo-assisted noise (Eq.\eqref{eq:super_poisson}).  
The photo-assisted noise becomes Poissonian only when the AC voltage consists of Lorentzian pulses, provided that the initial distribution is thermal and the zero-temperature limit remains within the validity domain of the UNEP approach \cite{ines_cond_mat,ines_photo_noise_PRB_2022}. This is however not the case for the weak backscattering noise within the TLL model—contrary to the claims of Ref.\cite{martin_sassetti_prl_2017}.\\
Another intriguing aspect of the FQHE is the expected anyonic statistics of fractional quasiparticles \cite{review_feldman}. Standard two-terminal Hong-Ou-Mandel setups are insufficient to reveal these statistics, but might require instead QPC sources driven by AC voltages within an "anyon collider" configuration \cite{glattli_levitons_physica_2017}, as illustrated in Fig.~\ref{fig:anyon_collider_AC}. A theoretical proposal in the DC regime, explicitly based on the TLL model \cite{fractional_statistics_theory_2016}, was successfully realized in pioneering experiments \cite{fractional_statistics_gwendal_science_2020,pierre_anyons_PRX_2023,fractional_statistics_gwendal_PRX_2023}. However, in these experiments as well, the measured DC current does not align with the typical power-law behavior. Identifying relations that reveal anyonic statistics independently of the specific model remains a major challenge. UNEP relations are promising in this direction; they have already provided a robust method to extract a key parameter $\lambda$ related to anyonic statistics (see Eq.\eqref{eq:anyon}) using finite-frequency admittance or noise \cite{ines_PRB_R_noise_2020}. \\
This paper is organized as follows: The next section provides a brief review of the underlying model and some universal relations within UNEP approach. We begin by considering an arbitrary initial non-equilibrium distribution, thus including the "anyon collider", followed by a discussion of an initial thermal distribution, where we recall the alternative characterization of minimal excitations. Sections three and four focus on DC and a careful study of photo-assisted transport in the quantum regime, particularly near resonant values of the D voltage. We also explore photo-conductance and photo-assisted noise for AC voltage profiles given by sine waves and Lorentzian pulses, addressing the challenges in characterizing minimal excitations and strategies to avoid resonant values of the DC voltage. Furthermore, we will highlight weaknesses in some prior studies. The last section addresses the "anyon collider" subject to AC voltages.
\section{Short review of the unifying non-equilibrium perturbative (UNEP) approach}
\label{sec:sec2}
The UNEP approach encompasses a broad class of physical systems, whose Hamiltonian consists of a time-independent part, $\mathcal{H}_0$, and a time-dependent component, $\mathcal{H}_B(t)$ determined by arbitrary complex function $p(t)$ and operator $B$\cite{ines_eugene,ines_cond_mat,ines_PRB_2019}:
\begin{subequations}
\begin{align}
	\label{eq:hamiltonian} 
	\mathcal{H}(t)&=\mathcal{H}_0 +\mathcal{H}_{B}(t),\\
\mathcal{H}_{B}(t)&=e^{-i\omega_Jt} {p}(t){B}+e^{i\omega_Jt}{p}^*(t){B}^{\dagger}
\end{align}
\end{subequations}
It requires however a weak observable derived from $\mathcal{H}_{B}(t)$, such as a charge or Josephson current, whose average and fluctuations under a constant $\omega_J$ or time-dependent drives encoded in $p(t)$ obey a set of universal relations. Sticking to charge transport, one deals with a charge current $I(t) = \partial_t Q(t)$ where $Q$ is a charge operator that gets shifted by a charge $e^*$ by the action of $B$, thus: $[B, Q] = e^* B$.\cite{note_IQHE} Consequently one has:
\begin{equation}
	\label{eq:current}
	I(t)\!=
		-i\frac{e^*}{\hbar}\left(e^{-i\omega_Jt}{{{p}}}(t)\;{B}- e^{i\omega_Jt}{p}^*(t) {B}^{\dagger}\right)\,.
\end{equation}  
 A crucial consequence of the UNEP approach is that for arbitrary $p(t)$ and $\rho_{neq}$, both the average and fluctuations of $I(t)$ at arbitrary frequencies are fully determined by two DC non-equilibrium observables that encode the signature of $\mathcal{H}_0$, $B$, and $\rho_{neq}$: the DC current average, $I(\omega_J)$, and the noise, $S(\omega_J)$. 
Let us recall some of the relations obtained at zero frequency and specified, for concreteness, to a periodic $p(t)$ with a frequency $\omega_{ph}$ \cite{ines_cond_mat, ines_eugene, ines_PRB_2019}:
\begin{equation}\label{eq:photo_current_periodic_zero}
{O}_{ph}(\omega_{J})= \sum _{l=-\infty}^{l=+\infty} P_l  O(\omega_{J}+l\omega_{ph})
\end{equation} 
 with $O=I,S$, and $O_{ph}$ the average of $O(t)$ over one period $2\pi/\omega_{ph}$. Here $P_l=|p_l|^2$, with $p_l=\omega_{ph}\int_{0}^{2\pi/\omega_{ph}} e^{i l\omega_{ph} t}p(t)dt /2\pi$. When $|p(t)| = 1$, $P_l$ represents the probability for the many-body eigenstates of $\mathcal{H}_0$ to exchange $l$ photons with the AC sources. The current $I(\omega_{J})$ is obtained from ${I}_{ph}(\omega_{J})$ in the limit where $p(t) = 1$, in which case $p_l=\delta_l$ (the Kronecker delta). 
The UNEP approach extends the Tien-Gordon theory of independent electron tunneling \cite{tien_gordon, tucker_rev, photo_reulet_belzig_PRB_2016} to account for strong correlations, but the side-band transmission picture must be reinterpreted in terms of many-body correlated states. For non-periodic $p(t)$, the discrete sum is replaced by an integral, suggesting an analogy to dynamical Coulomb blockade, with a classical environment simulated by time-dependent sources \cite{ines_cond_mat, ines_PRB_2019, mora2022anyonicexchangebeamsplitter}. \\
Let us also recall the Fourier transform of the current at a finite frequency, $m\omega_{ph}$, given that $p(t)$ is periodic:
\begin{eqnarray}\label{eq:TD_current_periodic_zero}
{I}_{m}(\omega_{J})&= &\left(i\frac{m\omega_{ph}}{2}-\epsilon\right)\int_{-\infty}^{\infty}\frac{d\omega}{(\omega+i\epsilon)(\omega-m\omega_{ph}-i\epsilon)}\nonumber\\&& \sum _{l=-\infty}^{l=+\infty} p^*_l p_{m+l}I(\omega_J+\omega+l\omega_{ph})\end{eqnarray} where $\epsilon$ is an infinitesimal real number (Eq. \eqref{eq:photo_current_periodic_zero} corresponds to $m= 0$). 
\subsection{Non-equilibrium initial distribution: Superpoissonian photo-assisted noise }
For any non-equilibrium distribution \( \rho_{neq} \), the universality of the super-Poissonian dc noise has been established in Ref.\cite{ines_cond_mat,ines_PRB_R_noise_2020}:  
\begin{equation}\label{eq:dc_superpoisson}
S(\omega_J) \geq e^* |I(\omega_J)|.
\end{equation}  
This inequality holds not only in the presence of a temperature gradient but also for non-equilibrium sources at zero temperature (this can be checked for the "anyon collider" from Eq.~\eqref{eq:anyon_collider_expressions}). Consequently, from Eq.~\eqref{eq:photo_current_periodic_zero}, one can deduce the universal nature of the super-Poissonian photo-assisted noise \cite{ines_cond_mat,ines_photo_noise_PRB_2022}:  
\begin{equation}\label{eq:super_poisson}
S_{ph}(\omega_J) \geq e^* |I_{ph}(\omega_J)|.
\end{equation}  
The same inequality applies for non-periodic \( p(t) \). This analogy with the DC case can be understood by interpreting \( p(t) \) as arising from a classical environment, which can be incorporated into the DC regime. 
\subsection{Equilibrium initial distribution: Minimal excitations}
Let us now consider an initial thermal distribution \( \rho_{eq} \propto e^{-\beta \mathcal{H}_0} \) at temperature \( T = 1/\beta \) and explicitly express the temperature dependence of all observables \( O \) through a frequency argument, \( \omega_T = k_B T / \hbar \), as \( O(\omega_J, \omega_T) \).
Then the above inequality gives a general alternative to the theorem by Levitov \textit{et al.} \cite{klich_levitov}, which states, in case \( p(t) = e^{-i\varphi(t)} \) and \( \varphi(t) \) periodic, that:
\begin{equation}\label{eq:levitov's_theorem}
S_{ph}(\omega_J, \omega_T) \geq S(\omega_J, \omega_T),
\end{equation}
for arbitrary transmission.  We have demonstrated the breakdown of this inequality in a nonlinear superconductor-insulator superconductor junction in Ref.\cite{ines_photo_noise_PRB_2022}, as well as within the TLL model in the present paper. In these two scenarios, the photo-assisted noise instead obeys the inequality in Eq.\eqref{eq:super_poisson}. In fact, Levitov's theorem applies only to conductors with a linear dc current, \( I(\omega_J,\omega_T) = G_{eq}(\omega_T) V \), where  $G_{eq}(\omega_T)$ is the linear conductance. In this case, the perturbative expression for the DC noise, which gives the lower bound in Eq.\eqref{eq:levitov's_theorem}, simplifies to  \( S(\omega_J, \omega_T) = e^* G_{eq}(\omega_T) V \coth\left( {\omega_J}/{2 \omega_T} \right) \). 

Let us consider again the UNEP approach, now specified to an initially thermalized non-linear conductor, for which it yields the non-equilibrium fluctuation-dissipation relation:
\begin{equation}\label{eq:noise_DC_zero}
S(\omega_J, \omega_T) = e^* \coth\left( \frac{\omega_J}{2 \omega_T} \right) I(\omega_J, \omega_T),
\end{equation}
which gives the Poissonian dc noise \( S_{neq}(\omega_J) = e^* I_{neq}(\omega_J) \) at $ \omega_T \ll \omega_J$ where we define \( O_{neq}(\omega_J) = O(\omega_J, \omega_T \ll \omega_J) \) for \( O = I, S \). It is important to note that Eq.~\eqref{eq:noise_DC_zero} extends beyond previous works \cite{kane_fisher_noise, levitov_reznikov}, which, for instance, do not account for interactions between the upper and lower edges.  

In the AC driven case, the photo-assisted noise $S_{ph}(\omega_J, \omega_T)$ is determined by both $p(t)$ and $S(\omega_J, \omega_T)$ through the sum of replica in Eq.\eqref{eq:photo_current_periodic_zero}. Adopting similarly to Levitov {\it et al} $p(t) = e^{-i\varphi(t)}$ and the zero temperature limit, we have shown that Poissonian photo-assisted noise, thus equality in Eq.\eqref{eq:super_poisson}, can only be achieved for:
\begin{equation}
\partial_t \varphi(t) = L(t) - n \omega_{ph},
\end{equation}
where \( L(t) \) is formed by a series of Lorentzian pulses with width \( W \) and a dc part \( \omega_J = n \omega_{ph} \) :
\begin{equation}\label{eq:V_lorentzian}
L(t)=\frac{n\omega_{ph}}{\pi}\sum_{k=-\infty}^{\infty}\frac{1} {1+(t-2\pi k/\omega_{ph})^2/W^2}.
\end{equation}As discussed in Appendix \ref{app:nonresonant}, $n$ does not necessarily correspond to the number of injected elementary charges per cycle in the FQHE, as was imposed in Ref.~\cite{klich_levitov} to reach equality in Eq.~\eqref{eq:levitov's_theorem} \cite{ines_photo_noise_PRB_2022}. This correspondence holds only for the Laughlin series. Furthermore, we will see that within the TLL model, backscattering noise under Lorentzian pulses remains super-Poissonian.

\section{Application of the UNEP approach to the FQHE}
\begin{figure}[htb]
    \centering
   \includegraphics[width=8cm]{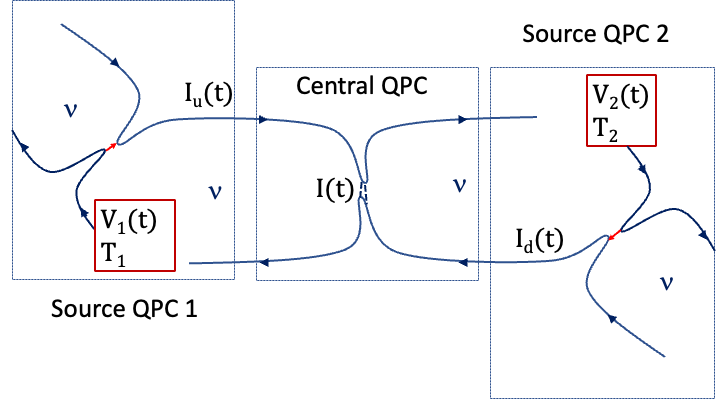}
    \caption{An "anyon collider" geometry where the source QPCs are driven by time-dependent voltages \( V_{1,2}(t) \) and may be at different temperatures \( T_{1,2} \). A common voltage profile with a time delay \( \tau \) between the two sources can be considered; however, precise control over the phase of \( p(t) \) remains challenging (see Eq.\eqref{eq:phiV0}). The UNEP approach remains applicable, asserting that time-dependent transport is formally governed by the non-equilibrium DC current and noise. If the TLL model is adopted, these quantities are given by Eq.~\eqref{eq:anyon_collider_expressions} \cite{fractional_statistics_theory_2016}.}
    \label{fig:anyon_collider_AC}
\end{figure}

We focus here on incompressible chiral Hall edge states at fractional filling factor \( \nu \) for which  \( \mathcal{H}_0 \)  in Eq.\eqref{eq:hamiltonian} can include inhomogeneous and finite-range interactions. Moreover,  \( \mathcal{H}_0 \) is not split into two terms, thus allows for mutual interactions between upper and lower edges. The initial non-equilibrium density matrix, $\rho_{neq}$, not specified, must commute with $\mathcal{H}_0$. The analysis carried in Appendix \ref{app:phi_relation} shows that controlling the phase of the complex function $p(t)$ through AC drives\( V_{1,2}(t) \) applied at sources is not straightforward, especially in the "anyon collider" configuration (see Fig.\ref{fig:anyon_collider_AC} and Section \ref{app:anyon_collider}).  Achieving this might be more feasible in a two-terminal geometry in the Laughlin states, $\nu = 1/(2k+1)$ with integer $k$.

It might be more feasible to control $p(t)$ locally via the gate voltage, thereby inducing a time-dependent modulation with $|p(t)|\neq 1$.  

Another challenge arises beyond the Laughlin series, where a single charge parameter $e^* = \nu e$ is present. For other filling factors $\nu$, multiple competing backscattering processes with different fractional charge values may exist, each characterized by a distinct scaling dimension $\delta_g$ that determines its relative weight. Labeling these processes by an index $g$, the perturbing Hamiltonian in Eq.~\eqref{eq:hamiltonian} becomes a superposition over $g$, where all parameters of the theory acquire subscripts: $e^*_g$, $B_g$, $\omega_{J,g}$, and $p_g(t)$. In this case, the corresponding UNEP relation must be applied to each individual process in Eq.~\eqref{eq:photo_current_periodic_zero}, with each photo-assisted and DC observables at $O=S,I$ acquiring also a subscript $g$. Omitting the dc arguments in the total photo-assisted ones (or one might write them implicitly through a vector formed by $\omega_{J,g})$, one has:
\begin{equation}\label{eq:superposition_photo_current_periodic_zero}
{O}_{ph}=\sum_g \sum _{l=-\infty}^{l=+\infty} P_{l,g}  O_g(\omega_{J,g}+l\omega_{ph}).
\end{equation} 
Therefore our current analysis regarding thermal contributions and constraints on the quantum regime remains applicable. In particular, the inequality in Eq.~\eqref{eq:super_poisson} now takes the form:
\begin{equation}\label{eq:superposition_super_poisson}
S_{ph}\geq \sum_g e^*_g |I_{ph,g}(\omega_{J,g})|.
\end{equation}
 If the dimensions $\delta_g$ of these processes—and consequently their weights—are comparable, it seems difficult to use this superposition still allows for the extraction of the individual charges $e^*_g$ involved. The methods we have developed within the UNEP framework are so far designed to identify a unique dominant charge \cite{ines_eugene, ines_cond_mat}.  

In practice, the application of these methods at $\nu=2/5$ in Ref.~\cite{glattli_photo} and $\nu=2/3$ in Ref.~\cite{ines_gwendal}, which provided evidence for $e^* = e/5$ and $e^* = e/3$, respectively, suggests that a single dominant backscattering charge $e^*$ prevails. However, we recall that the TLL model does not fully capture the details of these experiments.

Thus, we assume, for simplicity, that we can regroup backscattering processes with an effective backscattering amplitude $\Gamma_B$ and the lowest scaling dimension $\delta < 1$, where $\delta$ can be in general different from both $e^*/e$ and $\nu$. Notice that for co-propagating edges on each side, $\delta$ is identical to the statistical fractional phase $\theta/\pi$, but can be modified by non-universal inter-edge interactions or edge reconstruction. We can then write the backscattering operator, assumed to be localized at $x = 0$:
\begin{eqnarray}\label{eq:B}
   B^{\dagger} &=& \Gamma_B \Psi_{u}(0) \Psi_{d}^{\dagger}(0).
\end{eqnarray}
The quasiparticle fields are given by $\Psi_{\alpha}(x) = e^{i \Phi_{\alpha}(x)}/\sqrt{2\pi a}$ for $\alpha = u, d$, where $\Phi_u(x)$ and $\Phi_{d}(x)$ are upper and lower chiral bosonic fields which could be combinations of the bare edge bosonic fields and possibly of field boundary conditions (see Eq.\eqref{eq:phiV0}). The distance cutoff $a$ is on the order of the magnetic length. The reflection coefficient is given by $\mathcal{R} = \left[ \frac{|\Gamma_B|}{\hbar v_F} \right]^2$ if we adopt a single time cutoff $\tau_0 = a/v_F$ to which corresponds an UV frequency cutoff $2\pi/\tau_0$.

The expression for the photo-assisted noise in Eq.~\eqref{eq:photo_current_periodic_zero} applies both to a Hanbury Brown and Twiss setup, where the voltage is applied only in one reservoir, and to a Hong-Ou-Mandel-like setup, where \( \omega_J = 0 \) and \( p(t) = e^{i\left[\varphi(t + \tau) - \varphi(t)\right]} \), with \( \tau \) representing the time delay between the phase \( \varphi(t) \) in the left reservoir and \( \varphi(t + \tau) \) in the right reservoir. This is particularly relevant for the configuration of the "anyon collider" shown in Fig.~\ref{fig:anyon_collider_AC} (see Section~\ref{app:anyon_collider}), where the source reservoirs may be at different temperatures, potentially enabling the injection of fractional charges through the source QPCs. Notably, the time-independent Hamiltonian \( \mathcal{H}_0 \) differs in the "anyon collider" configuration compared to the two-terminal case, even though \( B \) in Eq.~\eqref{eq:B} retains a similar structure. However, as discussed in Appendix~\ref{app:phi_relation}, controlling the AC phase of \( p(t) \) induced at the central QPC is more challenging in this setup.

\section{Two-terminal configuration: DC transport}
In this section, we address the two-terminal geometry, and specify simultaneously to an equilibrium initial distribution, and to a frequency \( \omega_J \) that obeys a Josephson-type relation:  
\begin{equation}
\label{eq:joseph} \omega_J = \frac{e^* V}{\hbar}
\end{equation}with \( V \) is the dc voltage. 

In the stationary weak backscattering regime, the transmitted current through the QPC is given by  
\( I_0(\omega_J) - I(\omega_J, \omega_T) \),  where \( I_0(\omega_J) = \nu e^2 V / h = \nu e^2 \omega_J / (2\pi e^*) \). The backscattering current average in the TLL model is given by \cite{wen_photo_PRB_91}: 
\begin{equation}\label{eq:current_average}
I(\omega_J,\omega_T) = \frac{\hbar}{e^*} \omega_J G_{eq}(\omega_T) {F}\left( \frac{\omega_J}{2\omega_T} \right)\,.
\end{equation}
Here:
\begin{equation}
\label{eq:Fdefinition}
{F}(y)=\frac{\sinh{(y)}}{y}\,\frac{\left|\Gamma\left(\delta+i\frac{y}{\pi} \right)\right|^2}{\Gamma(\delta)^2},
\end{equation}
where $\Gamma$ is the Gamma function, and $G_{eq}(\omega_T)$ is the linear backscattering conductance in the equilibrium regime $\omega_J\ll \omega_T$:
\begin{equation}\label{eq:GT}
G_{eq}(\omega_T) = \frac{e^{*2} \mathcal{R}} h 
\frac{\Gamma(\delta)^2}{\Gamma(2\delta)}\,
\left[(2\pi)^2 \tilde{\omega}_T\right]^{2(\delta-1)}
\end{equation}
From now on, we adopt the notation $\tilde{\omega}=\tau_0 \omega/2\pi$ for all frequency scales $\omega$. The appearance of $e^{*2}$ is due to the renormalization of both $I$ and $\omega_J$ by $e^*$ (see Eqs.\eqref{eq:current},\eqref{eq:joseph}), assumed to be common for simplicity. 
The differential backscattering conductance can be obtained from Eq.\eqref{eq:current_average} :
\begin{equation}\label{eq:conductance}
G(\omega_J,\omega_T) = yG_{eq}(\omega_T)F(y)\left[\coth y-2 {\Im} \left(\Psi\left(\delta+{i y}/{\pi}\right)\right)/\pi\right],
\end{equation}
where $y=\omega_J/2\omega_T$. In the non-equilibrium limit, $\omega_T\ll\omega_J$, one gets:
\begin{equation}\label{eq:conductanceNE}
G_{neq}(\omega_J)\simeq \mathcal{R} \frac{e^{*2}}h \frac{(2\pi\tilde{\omega}_J)^{2(\delta-1)}} {\Gamma(2\delta-1)}.
\end{equation}
We also denote by:
\begin{equation}\label{eq:Ineq}
I_{neq}(\omega_J)\simeq \mathcal{R} \frac{e^{*}}{2\pi \tau_0}\frac{(2\pi\tilde{\omega}_J)^{2\delta-1}} {\Gamma(2\delta)}.
\end{equation}
Let us now discuss the validity of the weak backscattering regime in the two extreme non-equilibrium and equilibrium situations, $\omega_J\gg \omega_T$ or $\omega_J\ll \omega_T$, for which one has the power laws in Eqs.\eqref{eq:conductanceNE} or \eqref{eq:GT} respectively. For instance, one expects that $G_{eq}(\omega_T) \ll \nu e^2/h$ as long as $\omega_T\gg {\omega}_B$ where ${\omega}_B$ refers to a crossover frequency within the Bethe-Ansatz solution to the insulating strong backscattering regime \cite{fendley_prl_1}. But if two real numbers $a$ and $b$ verify $a\ll b$, and given a real $\gamma<0$, one has not necessarily $a^{\gamma}\gg b^{\gamma}$. Thus we prefer to impose an IR bound $\omega_{min}$ on $\omega_T$ through the condition $G_{eq}(\omega_{min})= \nu e^2/10 \, h$, a tenth the conductance in absence of backscattering:
\begin{eqnarray}
\tilde{\omega}_{min}=\tau_0\omega_{min}/2\pi=&=&\frac{1}{(2\pi)^2}\left[\frac{10 e^{*2}\Gamma(\delta)^2\mathcal{R}}{\Gamma(2\delta)\nu e^2} \,\right]^{\frac 1{2(1-\delta)}}.\label{eq:minimum_T}
\end{eqnarray}
A similar constraint on the non-equilibrium conductance in Eq.~\eqref{eq:GT} imposes a different IR bound on voltage, as the power-law behavior has different prefactors compared to Eq.~\eqref{eq:conductanceNE}. Nonetheless, both IR bounds are of the same order of magnitude. Thus, we avoid introducing an additional cutoff and simply impose \( \omega_J \geq \omega_{min} \) to ensure the non-equilibrium weak backscattering regime. Table~\ref{table:cutoffs} provides the values of \( \tilde{\omega}_{min} \) for two realistic values of \( \mathcal{R} \) and two scaling dimensions, \( \delta = 1/3 \) and \( \delta = 2/3 \).  
Notably, higher values of \( \delta \) may still be relevant to the FQHE (see, for instance, Ref.~\cite{kyrylo_FQHE_2024}), even when \( \nu = 1/3 = e^*/e \). They are also significant in the context of dynamical Coulomb blockade, as a QPC in the IQHE coupled to \( M \) edges \cite{ines_saleur,ines_pierre} can be mapped to the FQHE with \( \nu = \delta = M/(M+1) \). For example, \( M = 2 \) corresponds to \( \delta = 2/3 \) with \( e^*/e = \delta \), which modifies the value of \( \tilde{\omega}_{min} \).
\begin{table}[htb]
\begin{center}
\begin{tabular}{|c|c|c|}
\hline
$\mathcal{R}$& $\delta=1/3$, & $\delta=2/3$ \\
\hline
0.1 &  $6  \,\,\times 10^{-2}$&  $1.2\, \, \times 10^{-2}$\\
\hline
0.01 & $8\,\, \times 10^{-3}$&   $4\, \, \times 10^{-4}$\\
\hline
\end{tabular}
\end{center}
\caption{The values of the IR bound are given in units of the frequency cutoff, \( \tilde{\omega}_{min} = \tau_0 \omega_{min} / 2\pi \). Eq.~\eqref{eq:window} is imposed in both the equilibrium and non-equilibrium regimes to ensure that the backscattering conductance is below \( \nu e^2 / 10 \, h \), and that the low-energy effective theory remains valid. We have chosen \( \nu = 1/3 \) and \( e^* = e/3 \).}
\label{table:cutoffs}
\end{table}
Due to this lower bound, the non-equilibrium and equilibrium regimes where the power-law behavior is predicted may not be easily accessible. This is because, to ensure the validity of the TLL model, one must impose that any frequency scale  $\omega$  remains below the UV cutoff \( 2\pi/\tau_0 \), ensuring \( \tilde{\omega} \ll 1 \). The maximum allowed frequency scale depends on non-universal features, such as the cutoff procedure or the energy range of the linearized spectrum—for instance, whether the system is graphene or a 2D gas-based FQHE system. Assuming this scale is an order of magnitude lower, the window for observing power-law behavior is given by:
\begin{equation}\label{eq:window}
\tilde{\omega}_{min}\leq \tilde{\omega}\leq 0.1,
\end{equation}
where $\omega=\max(\omega_T,\omega_J)$. Such a window could disappear if \( \tilde{\omega}_{min} \) in Eq.~\eqref{eq:minimum_T} is too high, but it can be enlarged by reducing \( \tilde{\omega}_{min} \) through decreasing \( \mathcal{R} \) or increasing \( \delta \) or \( \nu \). For instance, both the non-equilibrium and equilibrium regimes are nearly unreachable for \( \mathcal{R} = 0.1 \). Only at \( \delta = 2/3 \) and \( \mathcal{R} = 0.01 \) can these regimes be explored over a range of three orders of magnitude (see Table~\ref{table:cutoffs}). This may explain the difficulty encountered when searching for evidence of power-law behavior in the FQHE over a few orders of magnitude in temperature or DC voltages, especially when \( \delta < 1/2 \). Thus, one might only access the intermediate regime, where the full expression for \( G \) in Eq.~\eqref{eq:conductance} must be used (see Appendix~\ref{app:intermediate}).

\section{Two-terminal setup: AC-driven quantum transport }

 As already mentioned, robust methods for determining the charge \( e^* \) can be derived from the UNEP relations without requiring knowledge of the underlying model for the edges. This presents a significant advantage for experimental works \cite{glattli_photo_2018, ines_gwendal}, which suggest a breakdown of the TLL model. When adopting the TLL framework and allowing \( p(t) \neq 1 \) in Eq.~\eqref{eq:hamiltonian}, we revisit the analysis of AC-driven quantum transport in the weak backscattering regime, where fractional charges are expected to tunnel.

Assuming initial thermalization, time-dependent transport within the UNEP approach is fully determined by a single function: the dc backscattering current characteristic \( I(\omega_J, \omega_T) \).
We focus on the relations obeyed by zero-frequency observables under an AC drive, such as the one in Eq.~\eqref{eq:photo_current_periodic_zero}, which holds for Hanbury-Brown and Twiss or Hong-Ou-Mandel setups.  The points of caution addressed here also concern high-frequency admittance and noise studied in Refs.~\cite{ines_cond_mat, ines_eugene, ines_degiovanni_2016, ines_PRB_R_noise_2020}.
\subsection{Backscattering photo-conductance}
 We begin our discussion with the photo-conductance, which is the derivative of \( I_{ph} \) in Eq.~\eqref{eq:photo_current_periodic_zero} with respect to \( V = \hbar \omega_J / e^* \):
\begin{equation}\label{eq:photo_G_periodic_zero}
{G}_{ph}(\omega_{J};\omega_{T})= \sum _{l=-\infty}^{l=+\infty} P_l  \,G(\omega_{J}+l\omega_{ph};\omega_{T}),\end{equation} 
where $G(\omega_J,\omega_{T})$ is given in Eq.\eqref{eq:conductance}. 
One can easily understand the origin of the above formula by expanding $p(t)$ in Eqs.\eqref{eq:hamiltonian}, \eqref{eq:B} into its Fourier components, so that one has a superposition of replica of backscattering processes with renormalized amplitudes $ p_l\Gamma_B$ and effective dc biases $\omega_J+l\omega_{ph}$:
\begin{equation}\label{eq:lambdal}
e^{i\omega_J t}p^*(t)B^{\dagger}= \sum_l p_l\Gamma_{B} e^{i(\omega_J+l\omega_{ph})t}\Psi_{u}(0) \Psi_{d}^{\dagger}(0).
\end{equation}
The validity of Eq.~\eqref{eq:photo_G_periodic_zero} requires that for each \( l \) contributing to the sum over replicas, the \( l \)-th argument on its right-hand side must be simultaneously below the UV cutoff to remain within the validity domain of the TLL model,  
\begin{equation}\label{eq:condition_l}
|\tilde{\omega}_J + l\tilde{\omega}_{ph}| \ll 1,
\end{equation}  
and within the weak backscattering validity domain defined by:
\begin{equation}
\label{eq:condition_replica_G}P_l|G(\omega_{J}+l\omega_{ph};\omega_{T})|\leq \nu e^2/10 \, h.
\end{equation}
A first point of caution arises when handling large arguments that do not satisfy Eq.~\eqref{eq:condition_l}, at which the TLL model breaks down. Formally, it is still legitimate to use Eq.~\eqref{eq:photo_G_periodic_zero}, derived independently of the underlying model, and thus the expression for the dc conductance at large arguments, which is expected to deviate from the TLL expression in Eq.~\eqref{eq:conductance}. The latter can be used only if the choice of $ \omega_J $, $ \omega_{ph} $, and $ p(t) $ ensures that the contribution of each $ l $ violating Eq.~\eqref{eq:condition_l} can be neglected. Such a contribution depends both on $ P_l $ \cite{Note_Pl} but also on the behavior of each replica of the observable one considers. For instance, when $ \delta > 1/2 $, and if we address $ I_{ph} $ and $ S_{ph} $ in the quantum regime, large $ l $-th terms exhibit a power law with a positive exponent $ 2\delta - 1 $, thus one needs even more decreasing $ P_l $. However, this is not the case for the photo-conductance in Eq.~\eqref{eq:photo_G_periodic_zero}, where the exponent is negative, $ 2(\delta - 1) < 0 $ (see Eq.~\eqref{eq:conductanceNE}).

A second point of caution arises as one might not reach the most interesting quantum regime given by \( \tilde{\omega}_{T} \ll \tilde{\omega}_{ph} \ll 1 \), in which case we refer to the short discussion in Appendix~\ref{app:intermediate}. One might expect that such a window is systematically opened alongside the dc stationary weak backscattering equilibrium or non-equilibrium regime, given by Eq.~\eqref{eq:condition_quantum_regime}. However, the window for \( \tilde{\omega}_{ph} \) is reduced due to the additional constraints in Eqs.~\eqref{eq:condition_l}, \eqref{eq:condition_replica_G}, and therefore depends on the dc voltage.

A third point of caution is specific to the neighborhood of resonant values of the dc voltage \( \omega_J = n \; \omega_{ph} \) with integer \( n \), defined by \( |\tilde{\omega}_J - n \tilde{\omega}_{ph}| \ll \tilde{\omega}_{T} \), where such a window becomes particularly narrow. In view of Eq.~\eqref{eq:lambdal}, the backscattering amplitude \( \lambda_{-n} \) experiences an effective voltage below \( \omega_T \), entering the equilibrium domain. As a result, the weak backscattering regime remains valid only for a sufficiently high temperature compared to a renormalized IR bound that depends on each \( n \) (see also Ref.~\cite{zaikinNov92} for a phase-slip Josephson junction):
\begin{equation}\label{eq:omegaBrenorm}
\tilde{\omega}_{min}(-n)=\, P_{-n}^{\frac{1}{2(1-\delta)}}\,\tilde{\omega}_{min}.
\end{equation}
where $\tilde{\omega}_{min}$ is given by Eq.\eqref{eq:minimum_T}. Notice that in case $|p(t)|=1$, thus for a constant tunneling amplitude, one has $P_{-n}<1$ so that $\tilde{\omega}_{min}(-n)< \tilde{\omega}_{min}$. 
This condition can also be deduced by setting $\omega_J = n\omega_{ph}$ in Eq.\eqref{eq:photo_G_periodic_zero}, which leads to an unavoidable temperature-dependent term even when $n\neq 0$ so that the dc voltage $|\omega_J = n\omega_{ph}| \gg \omega_T$ lies within the non-equilibrium domain:  
\begin{equation}\label{eq:Gph_res}
G_{ph}(n\omega_{ph};\omega_T)=P_{-n} G_{eq}(\omega_T)+\sum_{l\neq -n}P_l G_{neq}[(n+l)\omega_{ph}],\end{equation}
The condition in Eq.~\eqref{eq:condition_replica_G} at \( l = -n \) imposes precisely that \( \tilde{\omega}_T \geq \tilde{\omega}_{min}(-n) \) (see Eq.~\eqref{eq:omegaBrenorm}). Thus, contrary to the dc regime, the zero temperature limit can never be taken within the weak backscattering regime in the vicinity of $\omega_J = n\omega_{ph}$, as this drives the backscattering process with amplitude \( p_{-n}\Gamma_{B} \) (see Eq.~\eqref{eq:lambdal}) into the strong backscattering regime, and \( P_{-n} G_{eq}(\omega_T) \) would diverge.

\begin{figure}[ht]
 \begin{center}
\includegraphics[width=8cm]{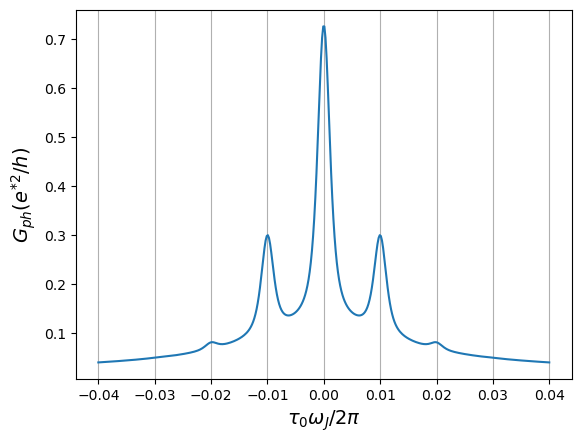} 
\end{center}
\caption{The differential photo-conductance given by Eq.~\eqref{eq:photo_G_periodic_zero}, using the full expression for the DC backscattering conductance \( G \) in Eq.~\eqref{eq:conductance}, and divided by $e^{*2} \mathcal{R}/(10 h)$. It exhibits peaks at \( \omega_J = 0, \pm \omega_{ph} \) which offer a method for determining the fractional charge. Here $\delta=2/3$, $\mathcal{R}=\tilde{\omega}_{ph}=\tau_0\omega_{ph}/2\pi=0.01$, and $\tilde{\omega}_T=\tau_0 k_B T/h=5\times 10^{-4}$. The ac voltage has a sine form with amplitude $e^*v_{ac}=\hbar \omega_{ph}$, thus $P_l=|J_l(1)|^2$ for integer $l$ with $J_l$ the modified Bessel functions.}
\label{Fig:Gph_2_3}
\end{figure}

The second term on the r.h.s. of Eq.\eqref{eq:Gph_res}  (the truncated sum) does not depend on temperature, as \( G_{neq} \) is given by Eq.~\eqref{eq:conductanceNE}, but it does not dominate the first term. In fact, since \( \omega_T \ll \omega_{ph} \), one has \( G_{neq}[(n+l)\omega_{ph}] < G_{eq}(\omega_T) \) for any \( l \neq -n \) (but not necessarily much smaller); this is because \( \delta < 1 \), so that \( \tilde{\omega}^{2(\delta-1)} \) is a decreasing function of \( \tilde{\omega} \). Therefore, we cannot neglect the equilibrium contribution to \( G_{ph} \) in Eq.~\eqref{eq:Gph_res}, which also holds for \( G_{ph}(\omega_J; \omega_T) \) at \( |\tilde{\omega}_J - n \tilde{\omega}_{ph}| \ll \tilde{\omega}_T \).
For these reasons, a proximity of the dc voltage to a resonance might shrink or even close the window for the quantum regime. One must have (see Eqs.\eqref{eq:condition_l},\eqref{eq:condition_replica_G}) :
\begin{equation}\label{eq:condition_quantum_regime}
\tilde{\omega}_{min}(-n)\leq\tilde{\omega}_{T}\ll \tilde{\omega}_{ph}\ll \min_l(|l+n|^{-1}),
\end{equation}
where the minimum is taken over all $l$ contributing to the sum of replica, which depend in particular on the behavior of $P_l$ and the power-law exponent. This yields $\omega_{ph}\gg {\omega}_{min}(-n)$, which is however insufficient for the validity of the non-equilibrium truncated sum in Eq.\eqref{eq:Gph_res} that imposes \begin{equation}\label{eq:non-equilibrium_condition}\omega_{min}(l)\leq |(n+l)\omega_{ph}|,\end{equation}
for each of the terms at \( l \neq -n \) (see Eq.~\eqref{eq:omegaBrenorm}). In case \( |p(t)| = 1 \) and $\partial_t\varphi(t)=e^*v_{ac}(\sin\omega_{ph}t )/\hbar$, $P_l\leq P_0$ for all $l$,  this last condition simplifies to $\omega_{min}(0)\leq \omega_{ph}$ independently on $n$. 
Let us further choose \( e^*v_{ac} = \hbar \omega_{ph} \), \( \delta = 1/3 \), and \( \mathcal{R} = 0.1 \). This gives \( \tilde{\omega}_{min}(0) = 0.04 \) (see Eq.~\eqref{eq:omegaBrenorm} and Table~\ref{table:cutoffs}). We can check that access to the weak backscattering regime near \( \omega_J = 0 \) or \( \omega_{ph} \) is not possible. In particular, we cannot address the Hong-Ou-Mandel setup, at least for a large range of delay times \( \tau \) (as will be explored in a separate publication).\\
 However, if we reduce the reflection coefficient further to $ \mathcal{R} = 0.01 $, its lowest realistic value, this allows us to access $ n = 0 $, and thus the Hong-Ou-Mandel setup, but not $ \omega_J = \omega_{ph} $, which is typically required to achieve one charge per cycle for the Laughlin series. Consequently, we cannot plot $ G_{ph} $ as a continuous function of the dc voltage (thus $ \omega_J $) at $ \delta = 1/3 $. To achieve this, we also need a higher scaling dimension, such as $ \delta = 2/3 $, to ensure a wider window for the quantum regime. Those are the values of $\mathcal{R},\delta$ chosen for all the curves plotted in this paper. 
Figure \ref{Fig:Gph_2_3} shows the differential photo-conductance in Eq.\eqref{eq:photo_G_periodic_zero} using the full expression of the differential DC conductance $G$ in Eq.\eqref{eq:conductance}. It has peaks at $\omega_J=0,\pm\omega_{ph}$ that provide a supplementary method to determine the fractional charge.
However, it is possible to consider the zero temperature limit at non-resonant values of the dc voltage, as discussed in Appendix \ref{app:zeroT}, where we provide a simplified but stronger version of the conditions in Eqs.\eqref{eq:condition_l}, \eqref{eq:condition_replica_G}.
\subsection{Photo-assisted backscattering current and noise}
Let us now express the photo-assisted current from Eq.~\eqref{eq:photo_current_periodic_zero} at resonant values. At $ l = -n $, the equilibrium term vanishes since $ I(0, \omega_T) = 0 $. However, reaching this zero-voltage limit still requires $ \omega_T \geq \omega_{min}(-n) $ (see Eq.~\eqref{eq:omegaBrenorm}) to remain within the weak backscattering regime. Under the same conditions ensuring the weak backscattering quantum regime discussed earlier for the photo-conductance, the photo-assisted current near $ n\omega_{ph} $, specifically for $ |\omega_{J}-n\omega_{ph}|\ll \omega_T $, can be expressed as:
\begin{eqnarray}\label{eq:Iph_res}
I_{ph}(\omega_J,\omega_T)&=&P_{-n}\frac{\hbar}{e^*}(\omega_{J}-n\omega_{ph})G_{eq}(\omega_T)+\nonumber\\&&\sum_{l\neq -n}P_l I_{neq}(\omega_J+l\omega_{ph}),
\end{eqnarray}
where the equilibrium conductance is given by Eq.\eqref{eq:conductance} while non-equilibrium dc current, intervening for $l \neq -n$, is given by Eq.\eqref{eq:Ineq}.
\begin{figure}[htb]
\begin{center}
\includegraphics[width=8cm]{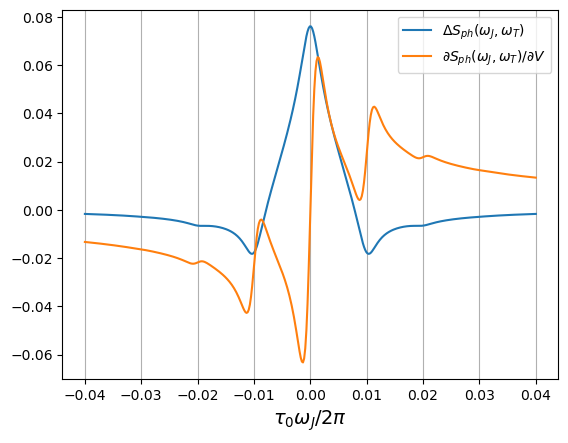}
\end{center}
\caption{Same parameters as in Fig.\ref{Fig:Gph_2_3}. The blue curve shows the excess photo-assisted backscattering noise, defined as the difference between the photo-assisted noise and the dc noise $S$ at the same dc voltage, $\Delta S_{ph}(\omega_J,\omega_T=5\times 10^{-4} \times 2\pi/\tau_0)$, divided by $e^{*2} \mathcal{R}/(10\tau_0)$. The orange curve represents its derivative with respect to the voltage $V$, divided by $e^{*3} \mathcal{R}/(10 h)$ (both are dimensionless).
We have used Eqs.\eqref{eq:photo_current_periodic_zero},\eqref{eq:current_average},\eqref{eq:noise_DC_zero}.}
\label{fig:excessS_2_3}
\end{figure}

The situation is different when one considers the finite frequency current $I_{m}(n\omega_{ph},\omega_T)$ in Eq.\eqref{eq:TD_current_periodic_zero} (we have added the temperature argument), which still contains an equilibrium term:
\begin{equation}\label{eq:FF_current_eq}I_{m}(n\omega_{ph},\omega_T)=f_{-n,m} \tilde{\omega}_T^{\; \; 2\delta-1}+\tilde{I}_{m}(n\omega_{ph}),
\end{equation}
where $f_{-n,m}$ is a complex number and $\tilde{I}_{m}(n\omega_{ph})$ is a non-equilibrium contribution that does not depend on temperature.

Now, we analyze the photo-assisted backscattering noise $ S_{ph}(\omega_{J};\omega_{T}) $, which follows the same relation as Eq.~\eqref{eq:photo_G_periodic_zero}, replacing $ G $ with $ S $, with the dc noise given by Eqs.~\eqref{eq:noise_DC_zero} and \eqref{eq:current_average}. The quantum regime is typically expected to suppress thermal fluctuations. However, we have shown that this regime is not always accessible near resonant values of the dc voltage. Even when it is reached, a thermal contribution remains unavoidable, as the noise retains a form similar to Eq.~\eqref{eq:Gph_res} in the vicinity of a resonance:
\begin{equation}\label{eq:Sph_res}
S_{ph}(n\omega_{ph};\omega_T) = P_{-n} S_{eq}(\omega_T)+e^*\sum_{l\neq -n}P_l |I_{neq}[(n+l)\omega_{ph}]|,
\end{equation}
which is subject to the same conditions as those on the photo-conductance. We have encountered a similar feature for the finite-frequency non-equilibrium noise in Ref.\cite{ines_degiovanni_2016}. The first contribution, the equilibrium dc noise \cite{ines_photo_noise_PRB_2022}, given by \( S_{eq}(\omega_T) = 2\omega_T G_{eq}(\omega_T) / \hbar \propto \tilde{\omega}_{T}^{2\delta-1} \), cannot be eliminated by any choice for the excess photo-assisted backscattering noise, as it depends on the ac voltage through \( P_{-n} \). The expression for \( I_{neq} \) is provided in Eq.~\eqref{eq:Ineq}, thus $|I_{neq}[(n+l)\omega_{ph}]|\propto \omega_{ph}^{2\delta-1}$. Since the exponent of the power law in both contributions is given by $2\delta - 1$, which is greater than $2\delta - 2$—the exponent governing the two terms in $G_{ph}$ in Eq.~\eqref{eq:Gph_res}—we expect that the spikes in $S_{ph}$ are less pronounced than those in $G_{ph}$.

\begin{figure}[htb]
\begin{center}
\includegraphics[width=8cm]{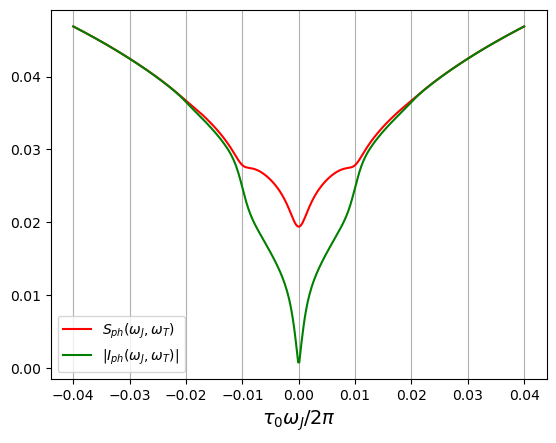}
\end{center}
\caption{The same parameters as in Fig.~\ref{Fig:Gph_2_3} are used. The photo-assisted noise is above the absolute value of the photo-assisted current, divided by \( e^* \mathcal{R} / (10\tau_0) \) and \( e^{*2} \mathcal{R} / (10\tau_0) \), respectively.}
\label{fig:photo-assisted noise_Iph_2_3}
\end{figure}
The plot of $\partial S_{ph}(\omega_J,\omega_T)/\partial V$ in Fig. \ref{fig:excessS_2_3} has peaks which are not located at the resonant values, thus is not convenient to access the fractional charge. 
We have also plotted the excess photo-assisted noise in the same figure, defined by subtracting the dc noise at the same dc voltage, $S(\omega_J,\omega_T)$: $\Delta S_{ph} = S_{ph} - S$. Here small cusps appear at $\pm \omega_{ph}$. As previously shown in the case of a SIS junction \cite{ines_photo_noise_PRB_2022}, it becomes evident that $\Delta S_{ph} < 0$ for all $|\omega_J|$ above a specific threshold value at which $S_{ph}$ equals $S$. This clear violation of Levitov's "theorem" \cite{klich_levitov} in Eq.\eqref{eq:levitov's_theorem} arises due to the nonlinearity of the dc current in the QPC. This further raises compelling questions about the applicability of the term "photo-assisted" noise, as it can, rather paradoxically, decrease when the QPC is irradiated with photons. We have also carefully plotted the photo-assisted backscattering noise alongside the absolute value of the backscattered photo-assisted current in Fig.\ref{fig:photo-assisted noise_Iph_2_3}, thus providing a solid verification of the alternative inequality in Eq.\eqref{eq:super_poisson}.
\subsection{Minimal excitations: some challenges}
\begin{figure}[ht]
\begin{center}
\includegraphics[width=8cm]{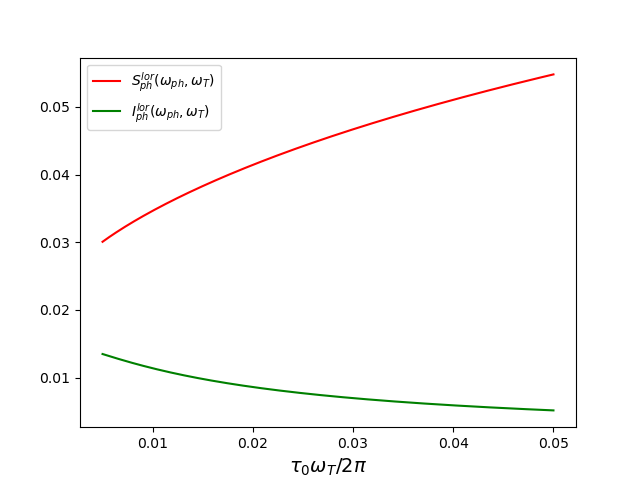}\end{center}
\caption{The photo-assisted backscattering current and noise as functions of temperature are shown for Lorentzian pulses with a width \( W \) given by \( W\omega_{ph} = 0.25 \), and with parameters \( \tilde{\omega}_J = \tilde{\omega}_{ph} = 0.01 \), \( \delta = 2/3 \), and \( \mathcal{R} = 0.01 \). The current and noise are divided by \( e^* \mathcal{R} / (10\tau_0) \) and \( e^{*2} \mathcal{R} / (10\tau_0) \), respectively. The photo-assisted backscattering noise remains super-Poissonian even in the quantum regime, due to the term \( P_{-1} S_{eq}(\omega_T) \) (see Eq.~\eqref{eq:super_poisson}).}
\label{fig:lorentzianphoto-assisted noise_Iph_2_3}
\end{figure}

We recall that if the zero temperature limit can be reached within the validity domain of the UNEP approach, equality in Eq.\eqref{eq:super_poisson}, thus Poissonian photo-assisted noise, can be reached only when $\partial_t\varphi(t) = L(t) - n\omega_{ph}$ where $L(t)$ is given by Eq.\eqref{eq:V_lorentzian}. We notice however that it is not guaranteed that one can control the shape of the pulses at the QPC level, particularly outside the Laughlin series, is addressed in Appendix \ref{app:phi_relation}. 
But assuming this is the case, and since the zero temperature limit is impossible to reach within the weak backscattering regime of the TLL model \cite{ines_photo_noise_PRB_2022}, the photo-assisted backscattering noise generated by these Lorentzian pulses (with the subscript $\text{lor}$) stays super-Poissonian. It still contains the equilibrium contribution \cite{note_spectroscopy} in Eq.\eqref{eq:Sph_res}, which cannot be neglected compared to the non-equilibrium contribution, identical to the absolute value of the photo-assisted current \cite{ines_cond_mat,ines_photo_noise_PRB_2022}:
\begin{equation}\label{eq:Slor}
    S_{ph}^{\text{lor}}(n\omega_{ph},\omega_T)=P_{-n}S_{eq}(\omega_T)+e^*|I_{ph}^{\text{lor}}(n\omega_{ph})|
\end{equation}
If we choose \( n > 0 \), this can be seen by comparing Eq.~\eqref{eq:Sph_res} to Eq.~\eqref{eq:Iph_res}, using the fact that \( P_l = 0 \) for all \( l < -n \). In the case where \( n = 1 \) and \( \delta = 2/3 \), we see that \( S_{ph}^{lor}(\omega_J = \omega_{ph}, \omega_T) \), plotted in Fig.~\ref{fig:lorentzianphoto-assisted noise_Iph_2_3} as a function of temperature (we impose \( \omega_T \geq \omega_{min}\)), is always above \( e^* |I_{ph}^{lor}(\omega_{ph})| \).
This challenges the claim that Lorentzian pulses minimize the backscattering photo-assisted noise or render it Poissonian, as thermal contributions remain significant. In fact, this is incompatible with achieving the zero-temperature limit, which requires well-chosen model parameters and a nonresonant value—either by applying an external dc voltage to shift away from $ n \omega_{ph} $ or by modifying the potential profile. In this case, the equilibrium contribution is eliminated, potentially leading to lower photo-assisted noise even without requiring the precise Lorentzian form in Eq.~\eqref{eq:V_lorentzian}—a feature that warrants further investigation. Alternatively, one can also achieve a nonresonant value by ensuring that the injected charge per cycle satisfies $ Q_{cycle} = N e^* $, with $ N $ an integer. As discussed in more detail in Appendix \ref{app:nonresonant}, this leads to a nonresonant dc voltage given by $ \omega_J = N\omega_{ph}/\nu $. One may then check whether the Lorentzian shape in Eq.~\eqref{eq:V_lorentzian} remains beneficial by replacing $ n $ with $ N/\nu $, or whether other pulse profiles could further reduce $ S_{ph}(N\omega_{ph}/\nu, \omega_T=0) $.  
Another way to deviate from the resonant values comes from a possible renormalisation of $\omega_J$ by the non-universal parameter $\lambda$, which depends on inter-edge interactions and edge reconstruction (see Eq.\eqref{eq:phiV0}).
\subsection{Comparison to prior works}
Let us now discuss a series of works on photo-assisted backscattering current and noise in light of our analysis. We select only some of them, but others performed within the same framework need to be checked. 
Let us first discuss Ref.~\cite{martin_sassetti_prl_2017} , which recovers,  by specifying to the TLL model, the inequality in Eq.~\eqref{eq:super_poisson} that we derived initially within the general framework of the UNEP approach~\cite{ines_cond_mat}. Nonetheless expressing directly the photo-assisted noise at zero temperature, which is again forbidden for the resonant values thus addressed. Our general criteria for minimal excitations, being those who achieve the Poissonian photo-assisted noise \cite{ines_cond_mat}, was incorrectly extended to the the photo-assisted backscattering noise, which stays in fact super-Poissonian\cite{ines_photo_noise_PRB_2022} (see Eq.~\eqref{eq:Slor}), with a temperature-dependent term that would diverge at zero temperature. 
Indeed, this work did not use the compact formula of the UNEP approach in Eq.~\eqref{eq:photo_current_periodic_zero}, but instead employed explicit expressions which make it more difficult to isolate, in the quantum regime and for resonant values of the dc voltage, the equilibrium term that is more transparent in our formula. It is precisely this term that leads to the divergence of the photo-assisted noise that the authors encountered in the zero temperature limit, and that cannot be canceled, contrary to their claim, by substracting $e^*\coth(\omega_J/2\omega_T)I_{ph}(\omega_J,\omega_T)$.\\
In other works, the divergence was not necessarily identified, and couldn't be anyway cancelled by other choices made for the excess photo-assisted backscattering noise, subtracting either the dc noise $S(\omega_J,\omega_T)$ (as we did in Fig.~\ref{fig:excessS_2_3}) or the equilibrium noise $S_{eq}(\omega_T)$. Such a divergence was artificially avoided in the photo-assisted current $I_{ph}(n\omega_{ph}, \omega_T)$, similarly to X. G. Wen, who incorrectly expressed it at zero temperature \cite{wen_photo_PRB_91}. Within the UNEP approach, using the expression of $I_{ph}(n\omega_{ph},\omega_T)$ in Eq.~\eqref{eq:photo_current_periodic_zero} , we have seen that the $-n$th term  vanishes only if we take care to keep a finite temperature above Eq.~\eqref{eq:omegaBrenorm}. In fact, it is not even sufficient to avoid divergences; only valid expressions within the weak backscattering regime can be used. Thus, one needs to impose a finite temperature $\omega_T \geq \omega_{min}(-n)$ in Eq.~\eqref{eq:omegaBrenorm} when addressing the neighborhood of $\omega_J = n\omega_{ph}$.\\
In fact, the reflection coefficient $\mathcal{R}$ has often been disregarded in those works, where it disappears through the renormalization of the observables, without addressing the UV cutoff or the IR bounds, which nevertheless depend on $\mathcal{R}$.  Since they were all carried out at $\delta = 1/3$, we choose the most favorable realistic value of the reflection coefficient, $\mathcal{R} = 10^{-2}$ to discuss further their results, and allowing us to lower the IR bound as much as possible, as $\tilde{\omega}_{min} = 8 \times 10^{-3}$ (see Table~\ref{table:cutoffs}).\\
In Ref.~\cite{crepieux_photo_proceedings_2005}, a finite temperature was introduced to avoid divergences in the differential photo-assisted noise at resonant values, but these results do not respect the UV cutoff for the validity of the TLL model. We have explicitly checked that we could obtain the same curve only if we tolerate excessively high arguments in the sum of replicas for which the TLL expression is not anymore justified.
The forbidden zero temperature limit was inappropriately taken at \( n = 1 \) in Ref.~\cite{martin_sassetti_levitons_heat_transport_PRB_2017} (see the figures 3, 5, 7) and \( n = 2 \) in Ref.~\cite{martin_two_leviton_FQHE_PRB_2024}. In addition, the series given for photo-assisted current and photo-assisted noise in Refs.~\cite{sassetti_99_photo, photo_crepieux, martin_sassetti_prl_2017}, based on the explicit LTL expression in Eq.~\eqref{eq:current_average}, are not justified when, for instance, Lorentzian pulses have too small width so that \( P_l \) do not decrease fast enough, thus terms with arguments exceeding the UV cutoff are beyond the TLL's validity domain.~\cite{lucas_ines_unpublished_2025}
Works on Hong-Ou-Mandel-like setups\cite{photo_crepieux, martin_sassetti_prl_2017, martin_sassetti_Hong-Ou-Mandel_heat_PRB_2018, martin_sassetti_crystallization_TD_PRB_2018, martin_sassetti_Hong-Ou-Mandel_multiple_levitons_EPJ_2018, martin_two_leviton_FQHE_PRB_2024} correspond to $\omega_J = 0$ or $n = 0$. We have shown that  they cannot be addressed for instance at $\mathcal{R} = 10^{-1}$ in the quantum regime, making the choice of $\mathcal{R} = 10^{-2}$ unavoidable. To simplify the discussion, we ignore the renormalization of the IR bound in Eq.~\eqref{eq:omegaBrenorm}, which depends on the ac voltage and time delay $\tau$, as it does not significantly affect the order of magnitude for $n = 0$ for most values of $\tau$. In Ref.~\cite{martin_sassetti_Hong-Ou-Mandel_heat_PRB_2018}, the authors considered temperatures that are too low compared to $\tilde{\omega}_{min} = 8 \times 10^{-3}$ (see their Figure 4, where $\tilde{\omega}_T = 10^{-6}/2\pi$). A similar issue may arise in Ref.~\cite{martin_sassetti_Hong-Ou-Mandel_multiple_levitons_EPJ_2018}.

Let us finally discuss works on the finite-frequency current average, or its time-dependent counterpart, shown here to contain also a temperature-dependent term even in the quantum regime through Eq.~\eqref{eq:FF_current_eq}. We raise some doubts about the crystallization of levitons in Ref.~\cite{martin_sassetti_crystallization_TD_PRB_2018}. This work is based on computing the time-dependent chiral current average, which is linearly related to the backscattering current through a relation derived in Ref.~\cite{ines_photo_noise_PRB_2022} (see also Ref.~\cite{dolcini_07}). Let us consider, for instance, Lorentzian pulses for the case $n=5$ addressed by the authors, for which the IR bound on temperature can be estimated through Eq.~\eqref{eq:omegaBrenorm} to be $\tilde{\omega}_{\text{min}}(-n=-5)=3\times 10^{-4}$, which is two orders of magnitude larger compared to the temperature chosen in that work, $\tilde{\omega}_T=10^{-5}/2\pi\simeq 10^{-6}$. This means that the carried weak backscattering analysis is not justified even for such a small reflection coefficient as $0.01$.

\section{AC-driven "anyon collider"}
\label{app:anyon_collider}
Let us now briefly discuss the case of the "anyon collider," where the two source QPCs are subject to AC voltages, injecting currents $I_u(t)$ and $I_d(t)$ into the upper and lower edges. In Fig.~\eqref{fig:anyon_collider_AC}, we also allow for two different temperatures to illustrate the extension of the UNEP approach to fully non-equilibrium setups. However, unlike the two-terminal case, the zero-temperature limit can now be taken, as the finite sum of the two DC components of the currents, $I_u$ and $I_d$, introduces an additional frequency scale that effectively replaces temperature, preventing the strong backscattering regime.  

To highlight the differences with the two-terminal configuration, let us first recall that the parameter $\omega_J$ in the UNEP approach, as defined in Eq.~\eqref{eq:hamiltonian}, is given by \cite{fractional_statistics_theory_2016,ines_PRB_R_noise_2020}:  
\begin{equation}
\label{eq:anyon} \omega_J = \frac{2\pi}{e^*}\sin(2\pi\lambda)(I_u-I_d).
\end{equation}
 The parameter $\lambda$ may correspond to the statistical anyonic phase, $\lambda = \theta/\pi$, but can also be modified by inter-edge interactions that induce charge fractionalization \cite{ines_epj} or by edge reconstruction \cite{Heiblum_edge_reconstruction_2018}. We have proposed a method to determine $\lambda$ through finite-frequency noise \cite{ines_bjorn_spin_hall_PRB_2020}, independently of the underlying model and thus of the power-law behavior predicted within the TLL framework. However, once the TLL model is adopted, explicit expressions for the DC average current and noise are given by \cite{fractional_statistics_theory_2016,ines_PRB_R_noise_2020}:  
\begin{eqnarray}
  I(\omega_J)&=&C'\mathcal{R}\sin \pi \delta \;\Im(\omega_++i\omega_J)^{2\delta-1},\nonumber\\
S(\omega_J)&=&C'e^*\mathcal{R}\cos \pi \delta \;\Re(\omega_++i\omega_J)^{2\delta-1},\label{eq:anyon_collider_expressions}
\end{eqnarray}
where $C'$ is a non-universal constant and $\omega_+=2\pi(\sin\pi\lambda)^2 (I_u+I_d)/e^*$.  
\begin{figure}[htb]
    \centering
    \includegraphics[width=8cm]{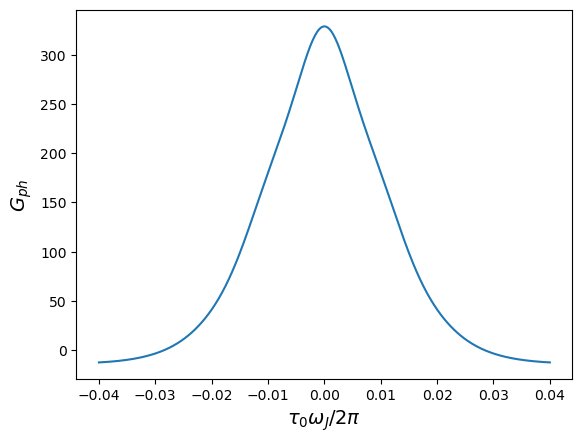}
        \caption{The photo-conductance in Eq.~\eqref{eq:photo_G_periodic_zero}, using the expression for the dc current in Eq.~\eqref{eq:anyon_collider_expressions} for the "anyon collider", is divided by $C' \mathcal{R} \sin(2\pi \lambda) (2\delta - 1)$. We take $\delta = \frac{1}{3}$ and assume a simplified sine AC voltage at the central QPC. We fix $\tilde{\omega}_+ = 0.01$.}
    \label{fig:anyon_collider_Gph}
\end{figure}
The UNEP relations can be exploited by injecting these expressions into Eq.~\eqref{eq:photo_current_periodic_zero} or its equivalent formula for non-periodic $p(t)$, thereby bypassing the difficulties encountered in the two-terminal geometry. The frequency $\omega_+$ introduces an additional scale that stabilizes the weak backscattering regime, allowing access to the zero-temperature limit in the presence of AC drives. Consequently, the UNEP approach formally permits consideration of a Hong-Ou-Mandel configuration, assuming Lorentzian pulses are applied at the QPC sources to inject single anyons.  
However, there are at least two additional challenges: the nature and charge of the excitations injected by the sources, and the intricate relationship between the phase $\varphi(t)$ of $p(t)$ and the potentials at the QPCs, as discussed in Appendix \ref{app:phi_relation}. Moreover, we have found that simply applying a sine form for $\varphi(t)$ to calculate the photo-conductance in Eq.~\eqref{eq:photo_G_periodic_zero} at $\delta=1/3$ does not yield any spikes at resonant values of $\omega_J$, as the finite scale $\omega_+$ smooths them out (see Fig.\ref{fig:anyon_collider_Gph}). Notice that with $\omega_+$ fixed, we can only explore values of $|\omega_J| \leq |\cot \pi \lambda| |\omega_+|$. All this complicates the use of photo-assisted transport as a reliable method for directly determining $\lambda$.

\section{Conclusion and Perspectives}
In this paper, we have investigated quantum transport in the context of the impurity problem within the TLL model, focusing on edge states in the FQHE at fractional filling factor $\nu$. We analyzed a QPC characterized by a scaling dimension $\delta$ and backscattering probability $\mathcal{R}$, carefully examining the validity of the weak backscattering regime, which depends on these parameters and the low-energy domain of the effective theory. Our findings indicate significant challenges in realizing the expected power-law behavior in both equilibrium and non-equilibrium DC regimes.\\
Under periodic time-dependent drives, we further explored the quantum weak backscattering regime, which proves even more difficult to achieve, particularly near resonant values of the DC voltage. Specifically, the zero-temperature limit cannot be reached within this regime, complicating the analysis of setups such as Hong-Ou-Mandel for certain parameter ranges. A thermal contribution to the photo-assisted backscattering conductance and shot noise persists, and while this contribution becomes weak at temperatures exceeding the infrared bound, it is never entirely negligible. However, by selecting $\delta = 2/3$ and $\mathcal{R} = 0.01$, we ensure access to the weak backscattering quantum regime, providing the first correct analysis of differential photo-conductance and photo-assisted noise. Notably, we show that the former is a more practical method for determining the fractional charge through spikes at resonant values of the dc voltage.\\
Due to the inherent thermal contribution within the TLL model, the Poissonian limit for backscattering photo-assisted noise cannot be reached using Lorentzian pulses \cite{ines_cond_mat, ines_photo_noise_PRB_2022}. We will propose a solution to this issue in a separate publication. Additionally, we qualitatively discussed how Lorentzian pulses may not retain their shape at the QPC, requiring experimental strategies to maintain their integrity. Imposing an integer number of fractional charges per cycle could help avoid resonances and mitigate thermal contributions. Future work should explore the role of non-periodic pulses, formally addressed by the UNEP approach \cite{ines_cond_mat,ines_photo_noise_PRB_2022,christian_non_periodic}.\\
We have also emphasized that several studies on photo-assisted transport in TLL liquids and Hong-Ou-Mandel-like setups have not sufficiently accounted for validity criteria, leading to questionable results, particularly when addressing the zero-temperature limit at resonant values.\\
Applying the UNEP approach to an AC-driven "anyon collider," we argued that the zero-temperature limit is permissible in this case, while also identifying several challenges for quantitative predictions.\\
More broadly, the UNEP approach applies to a variety of systems, including normal junctions, Josephson junctions, and dual phase-slip junctions strongly coupled to an ohmic environment \cite{zaikin_noise_perturbative_superconducting_wire_2017, glazman_arxiv24}. In the specific case of the TLL model, the scaling dimension $\delta$ for a small Josephson junction or a well-transmitting conductor connected to an ohmic environment with resistance $R$ is given by $1/\delta = R e^2/h$, as inferred from Ref.~\cite{grabert_ingold}, or $1/\delta = 1 + R e^2/h$, as established in Refs.~\cite{ines_saleur, ines_pierre}. A QPC connected to $M$ chiral edge states in the IQHE is therefore mapped to the FQHE with $\delta = M/(M+1)$. In these specific systems, at zero temperature, resonant values—such as the centers of Shapiro steps or their duals—fall outside the scope of the UNEP approach and will be addressed in future work.\\
Finally, the UNEP approach provides general relations governing current fluctuations at finite frequencies \cite{ines_cond_mat, ines_degiovanni_2016, ines_PRB_R_noise_2020}, which are particularly relevant for studying squeezing \cite{squeezing_reulet_PRL_2015} of emitted magnetoplasmons by a QPC. In the weak backscattering regime, adopting a TLL model forbids the zero-temperature limit in the FQHE, but not in the IQHE when the DC current remains linear ~\cite{squeezing_sasseti_NJP_2021, squeezing_gwendal_pascal_PRL_2023}.
Our current analysis can be extended to Fabry-P\'erot interferometers \cite{chamon_FPI_FQHE_PRB_1997,sukho_interferometers_FQHE_2012,halperin_2022}, where fractional statistics have been revealed in the DC regime \cite{manfra_braiding_anyons}, or Mach-Zehnder interferometers, particularly in light of recent experimental progress towards the injection of Lorentzian pulses in the IQHE \cite{interfero_gwendal} or in quantum wires \cite{interfero_seddik}. It may also be relevant to plasmonic cavities with a QPC capacitively coupled to ac-driven electrodes \cite{plasmon_cavity}.
A non-perturbative approach to AC-driven quantum transport seems challenging to address within the Bethe-Ansatz solution. However, for a dimension \( \delta = 1/2 \), relevant to \( \nu \neq \delta \), or a coherent conductor connected to a quantum of resistance \cite{ines_saleur}, this could be more feasible. So far, the DC regime has been explored to compute exact finite-frequency noise \cite{zamoum_12} and Hong-Ou-Mandel-like correlations \cite{giuliano_HBT_FQHE_PRB_2024}.
\section{Acknowledgments}  This work was supported by the ANR grant "QuSig4QuSense"
(ANR-21-CE47-0012). The authors thank Benoit Dou\c {c}ot, Lucas Mazzella and M. Seddik Ouacel for inspiring discussions. They thank G. M\'enard and G. Rebora for careful reading of the manuscript.
\appendix
\section{Intermediate regimes}
\label{app:intermediate}
 First, we discuss the stationary regime when temperatures and voltages lie between the two extreme equilibrium and non-equilibrium domains. It becomes more challenging to analytically express the domain where $G(\omega_J,\omega_T) \leq \nu e^2/10\, h$ (see Eq.~\eqref{eq:conductance}). For a fixed voltage (resp. temperature), this condition provides an infrared (IR) bound on temperature (resp. voltage), which depends on ${\omega}_J$ (resp. ${\omega}_T$), in addition to other non-universal parameters. We can show that $|G(\omega_J,\omega_T)|\leq G_{eq}(\omega_T)$ for all $\omega_J$. Thus, whenever ${\omega}_T \geq {\omega}_{\text{min}}$ in Eq.~\eqref{eq:minimum_T}, the weak backscattering regime remains valid for all values of ${\omega}_J$. Moreover, increasing ${\omega}_J$ lowers the IR bound on ${\omega}_T$. For instance, if we take ${\omega}_J=2\pi {\omega}_T$, the IR bound on temperature corresponds to approximately $\omega_{\text{min}}/10$ for both $\delta=1/3$ and $\delta=2/3$.  

For time-dependent drives, we consider the case where the ac quantum weak backscattering cannot be reached, for example, if $\tilde{\omega}_{ph} \leq \tilde{\omega}_T$. In this scenario, the first equilibrium term on the right-hand side of Eq.~\eqref{eq:Gph_res} remains, but in the second term, $G_{neq}[(n+l)\omega_{ph}]$ is replaced by the full expression $G[(n+l)\omega_{ph};\omega_T]$ given in Eq.~\eqref{eq:conductance}. The relatively small spacing between successive resonant values introduces additional equilibrium terms with their associated prefactors, leading to a higher renormalized value of $\tilde{\omega}_{\text{min}}$. In the extreme regime $\omega_{ph} \ll {\omega}_T$, all such terms sum up to $G_{ph}(n\omega_{ph};T)=G_{eq}(T)\sum_l P_l$. We recall that $\sum_l P_l=1$ whenever $|p(t)|=1$, in which case we recover the IR bound on temperature given in Eq.~\eqref{eq:minimum_T}.
\section{Zero temperature limit with AC drives}
\label{app:zeroT}
Here, we qualitatively discuss the extreme zero temperature limit. For resonant dc voltages, $\omega_J=n\omega_{ph}$, at least the term $p_{-n}\Gamma_B$ in Eq.~\eqref{eq:lambdal} undergoes a quantum transition in the dual strong backscattering regime. Although the UNEP approach framework allows one to treat strong backscattering perturbatively with respect to the dual amplitude, it is not justified to apply it to a mixture of resonant strong backscattering and nonresonant weak backscattering terms, which will be addressed in a separate publication. 

It is worth noting that the strong backscattering regime is generally less interesting, as no fractional charges are expected to tunnel. For non-resonant values of $\omega_J$, i.e., for all effective dc voltages satisfying $ |\tilde{\omega}_J +l\tilde{\omega}_{ph}|\gg \tilde{\omega}_T$, the equilibrium term is eliminated, and the system remains within the weak backscattering regime, provided the following condition holds (see Eqs.~\eqref{eq:condition_l}, \eqref{eq:omegaBrenorm}, and Table~\ref{table:cutoffs}).

\begin{equation}\label{eq:non-equilibriumconditions}
\tilde{\omega}_{min}(l)\leq |\tilde{\omega}_J +l\tilde{\omega}_{ph}|\ll 1 
\end{equation} This also requires, in turn, that $\tilde{\omega}_{min}(l)$ are low enough to open a window in which one can insert such effective dc voltages. We can simplify such conditions in a stringer fashion as follows by imposing   $\tilde{\omega}_{ph}\gg \max_l\tilde{\omega}_{min}(l)$. If we define $n$ so that $n\omega_{ph} <\omega_J <(q+1)\omega_{ph}$, it is then sufficient to require Eq.\eqref{eq:non-equilibriumconditions} only for $l=-q,-q-1$. Notice that only the lower bound at $l=-n$ was given in Refs.\cite{zaikin,glazman_arxiv24}.

\section{Control of AC-drives}
\label{app:phi_relation}
 Within the UNEP approach, the complex function \( p(t) \) in Eq.~\eqref{eq:hamiltonian} can encode simultaneous or separate time modulation of the backscattering amplitude and the voltages, whether random, non-periodic, or periodic. However, to keep the phase of $p(t)$ under control, the best but not obvious approach would be to induce it locally at the QPC. A gate is rather expected to induce an amplitude modulation, \( |p(t)| \neq 1 \).
If instead one imposes two time-dependent potentials \( V_{1}(t),V_{2}(t) \) on Gaussian or non-Gaussian QPC sources (as in the "anyon collider", see Fig.\ref{fig:anyon_collider_AC}), they inject total chiral currents \( I_{\alpha}(t) \) for \( \alpha = u, d \) in the upper and lower edges, which depend, respectively, in a linear or more nonlinear way on \( V_{j}(t) \). Then, \( p(t)=e^{-i\varphi(t)} \), where \( \varphi(t)=\varphi_u(t)-\varphi_d(t) \), can be determined through the equation of motion formalism for non-equilibrium bosonisation (pioneered in Ref.~\cite{ines_schulz,ines_epj}), with boundary conditions on edge \( j=u,d \) given by \( I_{u}(t), I_d(t) \). These are operators for QPCs sources \cite{out_of_equilibrium_bosonisation_eugene_PRL_2009}, illustrated in Fig.\ref{fig:anyon_collider_AC}. Thus, the phase of \( p(t) \) felt at the QPC obeys:
\begin{equation}
\omega_{J,\alpha}+\partial_t\varphi_{\alpha}(t)=\int_{-\infty}^t {\mathcal G}_{\alpha}(t-t') I_{\alpha}(t') dt'.  
\label{eq:phiV0}
\end{equation}  
where we have separated the upper and lower dc Josephson-type frequencies, such that $\omega_{J,u}-\omega_{J,d}=\omega_J$. ${\mathcal G}_{\alpha}$ is a Green's function which accounts for all edges on each side \( \alpha=u,d \) , describing the propagation of plasmonic collective modes over a distance \( L \) between the reservoir and the QPC position \cite{ines_ann}. In a two-terminal setup for simple fractions \( \nu \),  one expects \( {\mathcal G}_{u,d}(t)=\nu \delta(t\mp L/v) \). But more generally, ${\mathcal G}_{\alpha}$ may depend on the precise microscopic description of the edges and could be influenced by non-universal inter-edge interactions and edge reconstruction.  Indeed, in the "anyon collider" illustrated in Fig.\ref{fig:anyon_collider_AC}, the average of each $I_{\alpha}(t)$ for $\alpha=u,d$ can be obtained through a convolution with the backscattering current \cite{ines_photo_noise_PRB_2022} at each source QPC, given by Eq.~\eqref{eq:TD_current_periodic_zero} for each side.

\section{How to avoid resonant values of the dc voltage to mitigate the thermal contribution to the backscattering photo-transport.}
\label{app:nonresonant}
Usually, resonant values \( \omega_J = n\omega_{ph} \) ensure that an integer number of injected elementary charges \( ne \) is present per cycle. In a linear conductor where Levitov's theorem in Eq.~\eqref{eq:levitov's_theorem} holds, minimization of the photo-assisted noise was achieved while imposing that \( n \) is an integer. In case one has a linear dc current in the FQHE, such a minimization does not allow one to have an integer number of fractional charges \( e^* \) per cycle \cite{glattli_levitons_physica_2017}.
Within the UNEP approach for non-linear conductors, the resonance values arise from ensuring the Poissonian photo-assisted noise in the zero temperature limit \cite{ines_photo_noise_PRB_2022}. It is precisely when the function $\partial_t \varphi(t) + \omega_J$ is formed by Lorentzian pulses, where $\varphi(t)$ is the phase of $p(t) = e^{-i \varphi(t)}$ that intervenes in front of $B$ in Eq.\eqref{eq:hamiltonian}, that we get Eq.\eqref{eq:Slor}. But in that case, $n$ corresponds to the injected charges per cycle, $Q_{cycle}/e$ \cite{ines_photo_noise_PRB_2022}, only for simple fractions $\nu = 1/(2k + 1)$. This is because $Q_{cycle}$ is obtained by integrating the perfect linear edge current before the QPC, $I_0(t) = \nu e^2 V(t)/h$, so that, using the Josephson relationship $\omega_J = e^* V/\hbar$, one has:
\begin{equation}\label{eq:qcycle}Q_{cycle}/e=\frac{e}h \nu \int_{0}^{2\pi /\omega_{ph}}V(t)dt=\nu\frac{ e\omega_J}{e^*\omega_{ph}} \end{equation} 
Therefore, in the case of resonant values $\omega_J = n\omega_{ph}$, the injected charge per cycle is given by $Q_{cycle}/e = \nu n e / e^*$, which differs from $n$ outside the Laughlin series for which $e^* \neq \nu e$. Let us introduce the rational $r$ such that $\nu = r e^*/e$. Then $Q_{cycle}/e = r n$, which differs from $n$, and may be integer or rational depending on $r$. However, one could also start by giving a condition on $Q_{cycle}$, without regard to the minimization problem or "pseudo" Poissonian regime. For instance, if $r$ is rational, requiring that $Q_{cycle}/e$ be an integer leads to a non-resonant value of $\omega_J/\omega_{ph}$. But even if $r$ is integer, thus even for simple $\nu$ for which $r = 1$, one might also impose an integer number of fractional charges, $Q_{cycle} = N e^*$, so that $\omega_J/\omega_{ph} = N e^* / (r e)$ deviates generically from resonance, thus offering the possibility to reach the zero temperature limit as discussed before. Nonetheless, this does not ensure Eq.\eqref{eq:Slor} anymore, nor minimization of the photo-assisted noise in the case that the dc backscattering current was linear. Notice also that using the quantized conductance to evaluate the injected charge is not the only possibility, as one could estimate both in the reservoirs, for instance, which also gives another chance to achieve non-resonant values of $\omega_J/\omega_{ph}$.

This remark leads us to raise another important issue in case ac drives are not applied locally but from reservoirs. We have derived the non-trivial relation between the phase $\varphi(t)$ seen by the QPC and the injected chiral currents $I_{u,d}(t)$. In the "anyon collider", this leads to Eq.~\eqref{eq:anyon}, which deviates from the Josephson-type relation in Eq.~\eqref{eq:joseph}. However, the fractional charge in the latter could also deviate from $e^*$ in the two-terminal geometry we address here, being instead renormalized by the non-universal parameter $\lambda$. This would make it easier to move away from resonance, thus mitigating thermal contributions.



\begin{thebibliography}{10}
\bibitem{Note_Pl}
For instance, if we fix \( \omega_J = \omega_{ph} \), Lorentzian pulses in Eq.~\eqref{eq:V_lorentzian} with a width \( W \) large enough, such as \( W\omega_{ph} = 0.25 \), lead to rapidly decreasing \( P_l \) with \( l \). For a sine voltage with \( e^*V_{ac}/\hbar = \omega_{ph} \) and \( |\omega_J| \leq 3\omega_{ph} \) (our choice for the majority of the figures), one can generally ignore the terms with \( |l| \geq 3 \).
\bibitem{note_spectroscopy}
Notice that in case the complex function $p(t)$ is not known, one could access to its partial spectroscopy by varying $q$, as $(S_{ph}^{lor}-e^*| I_{ph}^{lor}(\omega_J,\omega_T=0)|)/S_{eq}(\omega_T)=P_{-q}$.
\end{thebibliography}

\end{document}